%% file: pi0eeg_prd.tex
\documentclass[twocolumn,amsmath,amssymb,showpacs,superscriptaddress]{revtex4}

\usepackage{graphicx}
\usepackage{dcolumn}
\usepackage{bm}

\def\pz{\pi^0}
\def\ee{e^+e^-}

\def\g{\gamma}
\def\ep{\epsilon}

\begin{document}

\preprint{APS/123-QED}

\title{Measurement of the Decay $K_L\to\pz\ee\g$}

\input{ktev99_author_list.tex}

\date{\today}

\begin{abstract}
We report on a new measurement of the branching ratio B($K_L\to\pz\ee\g$)
 using the KTeV detector. This analysis uses the full KTeV data set
collected from 1997 to 2000. 
We reconstruct 139 events over a background of 14, which results in 
B($K_L\to\pz\ee\g$) = $(1.62 \pm 0.14_{stat} \pm 0.09_{syst})
\times 10^{-8}$.
This result supersedes the earlier KTeV measurement of this branching ratio.
\end{abstract}

\pacs{13.20.Eb, 11.30.Er, 12.39.Fe, 13.40.Gp}
\maketitle

\section{\label{sec:intro}Introduction}

The decay $K_L\to\pz\ee\g$ can be used to study the low-energy dynamics
of neutral $K$ mesons. In particular, this decay is an important
check of Chiral Perturbation Theory (ChPT), which has been
used to describe kaon decays in which long distance effects dominate.
Up to $O(p^4)$ in chiral perturbation theory, there
are no free parameters and one predicts the branching ratio
to be approximately $1.0\times 10^{-8}$\cite{chpt}.
In the related decay $K_L\to\pz\g\g$, the $O(p^4)$ calculation
was found to underestimate the measured branching ratio by
a factor of three\cite{ref:na31a, ref:e731, ref:na31b, ref:ktev, ref:NA48}.
To match the $K_L\to\pz\g\g$ data it was
found necessary to extend the calculation 
to include $O(p^6)$ terms while introducing a free
parameter, $a_V$. This parameter characterizes the contributions from
vector meson exchange terms.\cite{av_ref}.
The addition of these effects to the
$K_L\to\pz\ee\g$ calculation results in an increase in
the branching ratio to $2.4\times 10^{-8}$, approximately
twice the $O(p^4)$ calculation. Our new branching ratio measurement
can distinguish between the $O(p^4)$ and $O(p^6)$ predictions.

Two previous experimental results have been reported
on this decay mode\cite{ggraham, keksearch, prl97}. 
The most recent measurement comes from the KTeV 1997 data
set and is based on 48 events with a background of
$3.6\pm 1.1$ events. That measurement yielded 
B($K_L\to\pz\ee\g$) = $(2.34\pm 0.35\pm 0.13)\times 10^{-8}$, 
where the rate was normalized to a now-obsolete value 
of B($K_L\to\pz\pz$). Using the latest measurement 
of B($K_L\to\pz\pz$)\cite{ref:2pi0BR, ref:PDG}, the 1997
measurement can be rescaled to $(2.17\pm 0.32\pm0.12)\times 10^{-8}$.
We report here on a new measurement of this mode from the
KTeV experiment.
For this analysis, we use improved techniques to reanalyze the 1997 data set 
and combine it with a new measurement from the 1999 data set.

The $K_L\to\pz\ee\g$ decay can also be used to help understand
the CP violating decay, $K_L\to\pz\ee$. The $K_L\to\pz\ee$ decay
contains both CP violating and CP conserving amplitudes. Since
the $K_L\to\pz\ee\g$ decay proceeds through a two photon intermediate
state, it can be used to determine the CP conserving components
in $K_L\to\pz\ee$, and thus allow one to determine the 
CP violating contribution in $K_L\to\pz\ee$. Also, because
the rate for $K_L\to\pz\ee\g$ is orders of magnitude higher
than the rate for $K_L\to\pz\ee$, a better understanding of
the $K_L\to\pz\ee\g$ decay will help to reduce the backgrounds
to the  $K_L\to\pz\ee$ decay.

\section{The KTeV Detector}

We collect $K_L\to\pz\ee\g$ events using the KTeV detector 
located at Fermilab. The data analyzed were taken during the
1997 and 1999 rare decay running periods and comprised
$2.9\times 10^{11}$ and 3.8 $\times 10^{11}$ kaon
decays, respectively. The KTeV experiment employed two different
configurations during its operation. The E799 configuration was 
used for this measurement and  was
optimized for reconstructing rare kaon decays.

In the KTeV experiment\cite{ref:detector} neutral kaons 
are produced in interactions of 800 GeV/$c$ protons
with a beryllium oxide target. The resulting
particles pass through a series of collimators to produce
two nearly parallel beams. The beams also pass through lead and
beryllium absorbers to reduce the fraction of photons and neutrons
in each beam. Charged particles are removed from the beams by sweeping
magnets located downstream of the collimators.
The vacuum decay volume begins approximately 94 meters downstream of the
target, far enough so that the majority of the $K_S$ mesons have
decayed, and extends to approximately 159 meters from the target. The
decay volume is surrounded by photon veto detectors 
that reject photons at angles greater than 100 milliradians. 

\begin{figure}
\includegraphics[width=8.5cm]{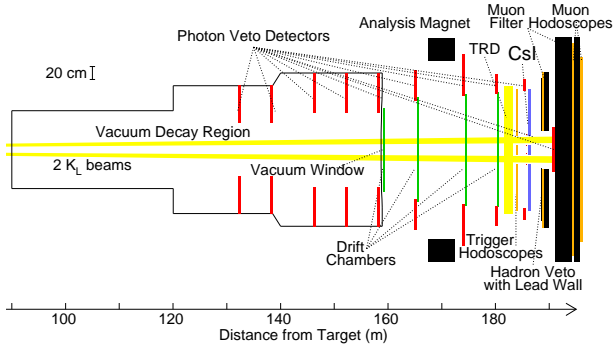}
\caption{\label{fig:ktevdet} Schematic of the KTeV detector.
}
\end{figure}

The most critical detector elements for this analysis are 
a charged particle
spectrometer and a pure CsI electromagnetic calorimeter\cite{csical}.
The KTeV spectrometer is used for reconstructing charged tracks. 
This spectrometer consists of four planes of drift chambers; two located
upstream and two downstream of an analyzing magnet with
a transverse momentum kick of 0.205 GeV/$c$. 
Each drift chamber contains four planes of wires; two to
measure the horizontal position and two to measure the 
vertical position with a precision of approximately 100$\mu$m.
During data taking in
1999 the momentum kick was reduced to 0.150 GeV/$c$ to increase the acceptance
for multi-track events.

The CsI calorimeter is composed of 3100 blocks in a 1.9 m by 1.9 m array.
The depth of the CsI calorimeter corresponds to
27 radiation lengths. Two 15 cm by 15 cm holes are located
near the center of the array for the passage of the two neutral beams.
For electrons with energies between 2 and 60 GeV,
the calorimeter energy resolution is below
1\% and the nonlinearity is less than 0.5\%. The position resolution
of the calorimeter is approximately 1 mm.
Downstream of the CsI calorimeter, there is a 10 cm lead wall, followed by
a hodoscope used to reject hadrons hitting the calorimeter.

The $K_L\to\pz\ee\g$ decays were required to satisfy 
certain trigger requirements in order to be recorded. 
In particular, activity in a set of hodoscopes upstream of the
CsI calorimeter had to be consistent with two tracks.
Also, we required the event have at least one hit in one of 
the two upstream drift chambers.  The event must
deposit more than approximately 25 GeV of total energy in the CsI 
calorimeter and no more than 0.5 GeV in the photon vetoes.  The 
event is vetoed if it deposits more than 2.5 Mips in the hodoscope
downstream of the calorimeter or more than
14 GeV in the vetos around the beam holes in the CsI
calorimeter.  The trigger includes a hardware cluster 
processor that counts the number of in-time calorimeter 
clusters of contiguous blocks of CsI  
with energies above 1 GeV\cite{hcc}. The total number of electromagnetic 
clusters in the CsI calorimeter is required to be 
greater than or equal to four at the trigger level.

After the events are read out, they must satisfy a software filter.
This filter requires that each event have two charged
tracks with a minimum of four clusters in the 
calorimeter. Each of the tracks must point to a cluster in the
calorimeter and be consistent with an electron hypothesis.
The trigger requirements also select $K_L\to\pz\pz_D$ 
events where one neutral pion undergoes Dalitz decay, $\pz\to\ee\g$ ($\pz_D$).
These events are
used for normalizing the $K_L\to\pz\ee\g$ events, since their
topology is very similar to that of our signal events. Because of
the similarity in topologies between the signal and normalization
modes, many systematic effects cancel.

\section{Event Reconstruction}
The offline analysis begins by requiring that 
each event have exactly two oppositely signed tracks and
five in-time clusters with energies greater than 2.0 GeV, where an
in-time cluster is one in which the cluster reconstructs to within
19 ns of the event time.
The two tracks are required to point to two
of the clusters and be consistent with a common decay vertex. 
From the three neutral clusters, we combine two
to form  the $\pz$ candidate. There are three possible 
combinations and we 
choose the combination that reconstructs closest to the
$\pz$ mass. 
Only events with a $\g\g$ invariant mass within 5 MeV/$c^2$ of the
nominal $\pz$ mass are accepted.
The neutral decay distance vertex is used to determine
the mass of the $\ee\g$ and $\ee\g\g\g$ combinations
because it improves their mass resolution; the $\ee$ tracks tend to
be close together leading to poorer vertex resolution.
The total kaon energy, determined from the sum of
cluster energies in the calorimeter, must lie between 30 and 210 GeV.

\begin{figure}
\includegraphics[width=8.5cm]{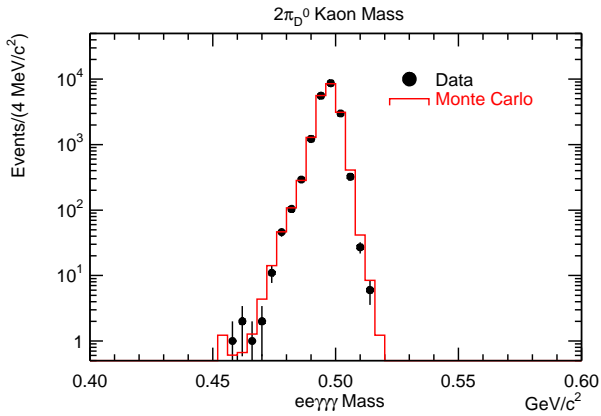}
\caption{\label{fig:2pi0} The $K_L\to\pz\pz_D$ invariant mass distribution
for data (dots) and $K_L\to\pz\pz_D$ Monte Carlo (solid histogram).
}
\end{figure}

To ensure that the two tracks are electrons,
the reconstructed energy in the calorimeter divided by the
momentum determined by the spectrometer ($E/p$) of each track must be between
0.95 and 1.05. Backgrounds from $K_s$ decays and misreconstructed
kaons can be reduced by
requiring the decay vertex to reconstruct
between 98 and 157 meters downstream of the target,
and the transverse momentum squared ($p_T^2$) for the event
to be less than 0.003 (GeV/c)$^2$.
The invariant mass for $K_L\to\pz\pz_D$ events is shown 
in Figure~\ref{fig:2pi0}. The data and our
Monte Carlo simulation agree quite well.

\section{Backgrounds to $K_L\to\pz\ee\g$}

After applying the above selection criteria, the remaining
backgrounds consist mainly of $K_L\to\pz\pz_D$ and
$K_L\to\pz\pz\pz_D$ decays.
The $K_L\to\pz\pz_D$ decays are more readily
removed because the invariant masses of the $\ee\g$ and $\g\g$ combinations
reconstruct around  the mass of the $\pz$. 
The majority of the $K_L\to\pz\pz_D$ events are removed
by requiring the reconstructed $\ee\g$ mass of the
best combination to be less
than 0.110 GeV/$c^2$ or greater than 0.155 GeV/$c^2$. However, when
the wrong $\g\g$ combination is chosen, a restriction on the
invariant masses is ineffective at reducing the background.
$K_L\to\pz\pz_D$ events can also contribute to the background
if one of the final state particles is lost and is replaced by
activity in the detector that can mimic a final state particle.
The $K_L\to\pz\pz\pz_D$ events
are more difficult to remove because we cannot use the same
mass constraints as in the $K_L\to\pz\pz_D$ case. However, kinematic
and cluster shape variables have been developed to
help to reduce the background to a manageable level.

To remove misreconstructed $K_L\to\pz\pz_D$ decays, we 
consider the two other possible $\g\g$ combinations.
We take advantage of the correlations between the
$m_{\g\g}$ and $m_{\ee\g}$ distributions for these 
two combinations, forming a neural net from four
variables. These four input variables are 
the reconstracted invariant $\g\g$ and $\ee\g$ masses
for each of the two remaining combinations.
The neural net employed
sixteen hidden nodes and was tuned on a sample of
$2\pz_D$ and $\pz\ee\g$ Monte Carlo. The output from the
neural net ranges between zero and one. We
rejected events where the neural net value was less than 0.5.

\begin{figure}
\includegraphics[width=8.5cm]{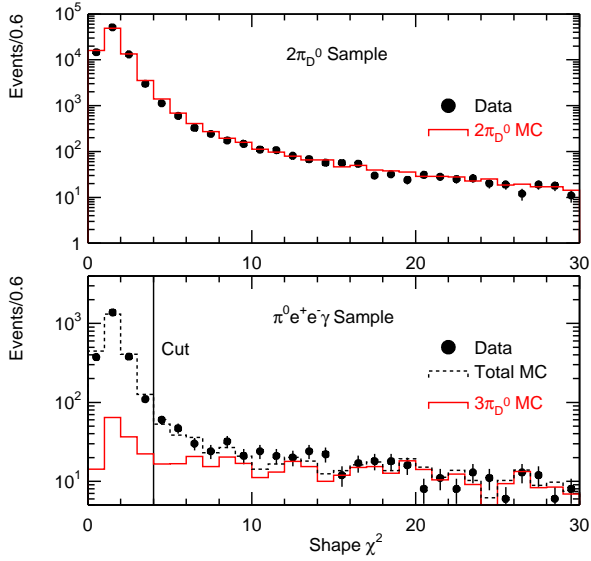}
\caption{\label{fig:fuse3x3} The photon shape variable shape $\chi^2$.
The top plot shows the shape $\chi^2$ variable for $K_L\to\pz\pz_D$
with the dots representing the data and the solid histogram
the Monte Carlo. In the bottom histogram the dots are the data
after removing the $K_L\to\pz\pz_D$ events, while the solid
histogram shows the $K_L\to\pz\pz\pz_D$ Monte Carlo. The dashed
histogram represents the sum of the signal and background Monte Carlo
samples.}
\end{figure}

Backgrounds from $K_L\to\pz\pz\pz_D$ come from two broad classes of events:
events with missing photons and those with one or more photons that
overlap or fuse together in the CsI calorimeter.
For events with missing photons, we use the photon vetoes to significantly
reduce the amount of background. We require the maximum energy in
any photon veto to be less than 0.1 GeV. To reduce backgrounds
from events with overlapping photons, we examined the calorimeter energies in
a $3\times3$ array of crystals centered around the
highest energy crystal of the cluster. For reference, a $3\times3$ array
corresponds to approximately one Moli\`ere radius.  We compared these
energies to energies from an ideal cluster shape and
calculated  a photon shape $\chi^2$ variable.
This variable is shown in Figure~\ref{fig:fuse3x3}.
As can be seen, for the normalization mode, there is good
agreement in this variable between the data and the Monte
Carlo simulation. For the signal events, the background
from $K_L\to\pz\pz\pz_D$ events is significantly reduced
by requiring a small value of shape $\chi^2$. We require 
 shape $\chi^2$ $<$ 4.

\begin{figure}
\includegraphics[width=8.5cm]{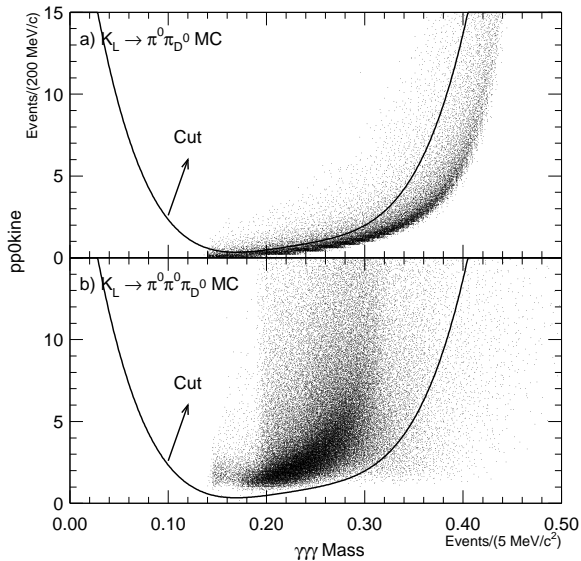}
\caption{\label{fig:pp0kin_vs_mggg} The pp0kine variable plotted versus
the $\g\g\g$ invariant mass for a) $K_L\to\pz\pz_D$ and
b) $K_L\to\pz\pz\pz_D$ Monte Carlo events. The dark line represents the
cut indicated in the text.}
\end{figure}

Kaon decays with missing
photons will also exhibit a significant amount of 
missing energy when boosted to the center-of-mass.
We take advantage of this effect by calculating the
longitudinal missing momentum in the center-of-mass
(pp0kine). In the pp0kine versus $m_{\g\g\g}$ plane,
the signal events are well-separated from the $K_L\to\pz\pz\pz_D$
background.
We define a two dimensional cut by employing the
following
fourth-order polynomial:
\begin{eqnarray*}
  \mbox{pp0kine}_{max} &=& A + B*(m_{\g\g\g}-x_0) + C*(m_{\g\g\g}-x_0)^2 \\
                      &+&
                        D*(m_{\g\g\g}-x_0)^3 +  E*(m_{\g\g\g}-x_0)^4.
\end{eqnarray*}
where A = 3.9, B=-112.8, C=1256.6, D=-5861.8, E=10506.0 and
$x_0 = 8.326\times 10^{-2}$. The values of these parameters were
chosen to maximize the signal-to-background ratio. Events with values of
pp0kine greater than this value were rejected. This cut
is shown in Fig.~\ref{fig:pp0kin_vs_mggg}.

\begin{figure}
\includegraphics[width=8.5cm]{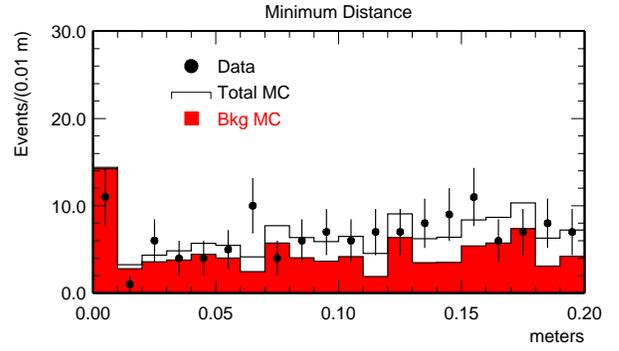}
\caption{\label{fig:mindist} The minimum distance between
the projection of the upstream electron track segment to the
CsI calorimeter and any photon candidate. The dots represent the
data and the solid histogram the sum of the background and signal
Monte Carlo. The enhancement at low values of the minimum distance
is the result of bremsstrahlung production.}
\end{figure}

Both $K_L\to\pz\pz_D$ and $K_L\to\pz\pz\pz_D$ events can contribute
to the final data sample if one of the electrons undergoes
bremsstrahlung radiation. To reduce this background, we
calculate the minimum distance 
between the projection of the
upstream segment of each electron to the CsI calorimeter
and any photon cluster. Bremsstrahlung photons tend to
have a small minimum distance. As shown in Fig.~\ref{fig:mindist},
backgrounds from $K_L\to\pz\pz_D$ and $K_L\to\pz\pz\pz_D$
occur at small photon-to-track distances. To reduce this
background, we require the minimum distance to be greater than 1.25 cm.

Backgrounds due to external conversions of photons are negligible
since we require the decay vertex to lie within the vacuum
decay region, upstream of the vacuum window. The requirement
that the neutral and charged vertex be consistent with each
other further reduces the probability of external conversions
contributing to the background since external conversions can only
result from interactions with material downstream of the
vacuum window.

\section{Results}
\begin{figure}
\includegraphics[width=8.5cm]{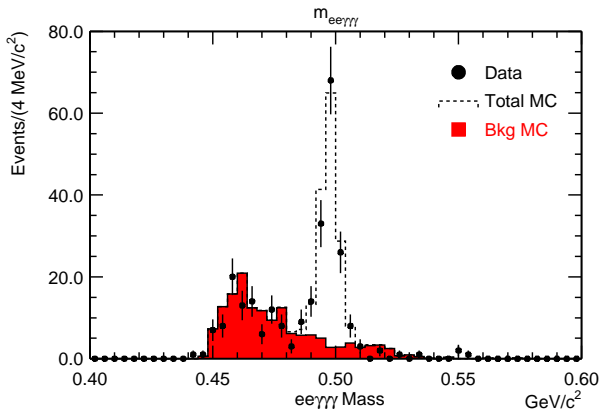}
\caption{\label{fig:meeggg} The $\ee\g\g\g$ invariant mass for events
passing all selection criteria. The dots represent the data, while
the dashed histogram represents the sum of the signal and background
Monte Carlo. The background Monte Carlo is indicated by the
shaded histogram.}
\end{figure}

After making these final selection criteria,
we find the 
$\ee\g\g\g$ mass distributions shown in Figure~\ref{fig:meeggg}. A clear
peak at the kaon mass is seen, while the background is well-described by
the sum of the $2\pz_D$ and $3\pz_D$ background Monte Carlo samples. 
We find a total of 139 candidate events with an estimated background
of $14.4\pm2.5$ events.

The $K_L\to\pz\ee\g$ branching 
fraction is determined from the following expression:
\begin{eqnarray*}
   B &=& (N_{\pz\ee\g}/N_{2\pz})\times (\ep_{2\pz}/\ep_{\pz\ee\g}) \\
      &\times& B(K_L\to\pz\pz) \times B(\pz\to\ee\g) \times 2.
\end{eqnarray*}
Here, $N_{\pz\ee\g}$ represents the number of signal candidates, while
$N_{2\pz}$ represents the number of normalization events.
The number of
$K_L\to\pz\pz_D$ candidates is determined by inverting the cut
against $K_L\to\pz\pz_D$ events and counting the number of events in
the kaon mass region from 0.490 to 0.510. In
the above expresssion $\ep_{2\pz}$ and $\ep_{\pz\ee\g}$ correspond 
to the reconstructed
$K_L\to 2\pz$ and $K_L\to\pz\ee\g$ acceptances, respectively.
The factor of two occurs because there are two $\pz$ in each $K_L\to\pz\pz_D$
event. In the previous analysis, the value of
B($K_L\to\pz\pz$) used was $(9.36\pm 0.2)\times 10^{-4}$.
We are now using the
most recent determination of B($K_L\to\pz\pz$) = $(8.69\pm 0.08)\times 10^{-4}$. 
The value of B($\pz\to\ee\g$)
used in both analyses is $(1.198\pm 0.032)\times 10^{-2}$.

The acceptance for $K_L\to2\pz_D$ events is 0.51\% in the 1997 data set
and 0.61\% in the 1999 data set. The difference between the acceptances
in the two data sets arises from the different magnetic fields used during
the 1997 and 1999 runs. We find 31,286 $K_L\to\pz\pz_D$ events in the 1997
data and 49,159 events in the 1999 data. This corresponds to a
kaon flux of $2.88\times 10^{11}$ and $3.78\times 10^{11}$
decays in the 1997 and 1999 data
sets, respectively.
The $\pz\ee\g$ acceptances are 0.90\% and 1.02\% for the
1997 and 1999 data sets, respectively. These values are
shown in Table~\ref{table:numbers}.
Using the numbers above, we obtain:
\begin{eqnarray*}
    \mbox{B}(K_L\to\pz\ee\g)&=& (1.49\pm0.22)\times 10^{-8} 
            \qquad \mbox{(1997)} \\
    \mbox{B}(K_L\to\pz\ee\g)&=& (1.77\pm0.18)\times 10^{-8} 
            \qquad \mbox{(1999)} \\
\end{eqnarray*}
The difference between the result in Ref. \cite{prl97}
and our new 1997 measurement
of the $K_L\to\pz\ee\g$ branching ratio can be mainly attributed to
the different values of $a_V$ used in determining the acceptance.
The previous measurement used $a_V=-0.96$ while our new
analysis uses $a_V=-0.46$\cite{ref:NA48}. After
accounting for this acceptance effect, we find that the two
analyses are consistent with each other. Our new measurement 
supersedes the previous KTeV measurement.

\begin{table}[tbh]
  \centerline{
    \begin{tabular}{l c c} \hline\hline
      Value  &  1997  & 1999\\ \hline 
      Events in Data           &  47      & 92   \\
      Background Events        &  2.7     & 11.7 \\
      Normalization Events     & 31,286   & 49,159 \\
      Signal Acceptance        &  0.91\%  & 1.03\% \\
      Normalization Acceptance & 0.51\%   & 0.61\% \\
      \hline\hline
    \end{tabular}
  }
  \caption{Values used in branching ratio calculation.}
  \label{table:numbers}
\end{table}

\section{Systematic Uncertainties}

The largest systematic uncertainty results from the limited 
statistics in our background Monte Carlo sample. In total we generated
approximately twice the statistics of the $K_L\to\pz\pz\pz_D$ 
data sample, and approximately three times the statistics of
the normalization and signal modes. This required generating about
six billion Monte Carlo events. The next largest
systematic uncertainty arises from the $K_L$ and $\pz\to\ee\g$
branching ratios.  The remaining 
effects can be  broken down into two main classes:
those that affect the background level and those that affect
the signal or normalization acceptance.
The signal acceptance has a dependence upon the
value of $a_V$. This dependence can be characterized by
$B = 1.484 + 0.215*a_V + 0.1.170 *a_V^2 + 0.417*a_V^3$.
We varied
the value of $a_V$ between -0.41 and -0.51\cite{ref:NA48},
and found the acceptance changed by approximately 2.0\%.
The uncertainty in the background contributes approximately
0.5\% to the total systematic uncertainty, and the remaining
acceptance effects including the effects of apertures and
cuts contribute about 0.6\% to the total
systematic error. All of the systematic errors are
listed in 
Table~\ref{table:systematics}.

\begin{table}[tbh]
  \centerline{
    \begin{tabular}{l c} \hline\hline
      Systematic & Error (\%)\\ \hline 
      MC Statistics                 &  4.2   \\
      $K_L\to\pz\pz$ and $\pz_D$ BR &  2.8  \\
      $a_V$ dependence              &  2.0   \\
      Signal acceptance             &  0.6   \\ 
      $3\pz_D$ and $2\pz_D$ background  &  0.5   \\ \hline
      Total                         &  5.2   \\ \hline
      \hline
    \end{tabular}
  }
  \caption{Systematic uncertainties in percent.}
  \label{table:systematics}
\end{table}

To obtain the final result, we took the weighted average of the
1997 and 1999 numbers, where we weighted by the statistical
error. The systematic studies were done on the combined 
1997 and 1999 analyses to take into account any correlations.
Including the uncertainties due to the systematic effects, we
find the following result:
B($K_L\to\pz\ee\g$) = $(1.62\pm 0.14_{stat} \pm 0.09_{syst})
\times 10^{-8}$.

The relatively small value for the 
$K_L\to\pz\ee\g$ branching ratio means that this decay will not
serve as a significant source of background to $K_L\to\pz\ee$. 
The $\pz\ee$ mass distribution from $K_L\to\pz\ee\g$ decays peaks
around 0.250 GeV/$c^2$, far enough away from the kaon mass that
only a small fraction of the decays pose any risk of reconstructing
in the $K_L\to\pz\ee$ signal region.

\begin{figure}
\includegraphics[width=8.5cm]{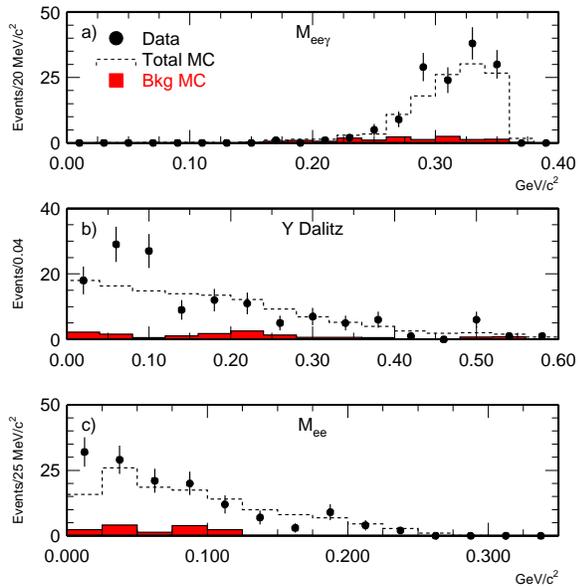}
\caption{\label{fig:dalitzvars} The Dalitz variables used
in the fit for $a_V$: a) $m_{\ee\g}$, b) $Y$ and c) $m_{\ee}$.
The dots are the data. The dashed histogram is the sum of the background
and signal Monte Carlo, and the filled histogram is the background
Monte Carlo.}
\end{figure}

To determine the value of $a_V$ from our data, we
performed a maximum likelihood fit to the three
Dalitz parameters $Z = \frac{m_{eeg}^2}{m_K^2}$,
$Q = \frac{m_{ee}^2}{m_K^2}$, and $Y = \frac{E_{\gamma}-E_{ee}}{m_K}$.
The variables
$E_{\gamma}$ and $E_{ee}$ are the energies of the photon and the $e^+e^-$ pair
in the kaon CM frame, respectively. 
The value of  $a_V$ that we obtained is $-0.76\pm 0.16\pm 0.07$. The
major systematics are the background level and  the sensitivity to
the selection criteria.
Our value for $a_V$ is consistent with the recent published values but
our errors are significantly larger.
The distributions of the Dalitz variables are shown in Fig.~\ref{fig:dalitzvars}.
We find  good agreement between the data and the Monte Carlo generated
with the central $a_V$ value. The shapes of the
$m_{\ee\g}$ and $Y$ variables are similar to those seen
in $K_L\to\pz\g\g$ decay. In principle the $\ee$ mass distribution
should rise sharply near threshold. However, due to the detector acceptance this
peaking is suppressed.

In Ref.\cite{chpt} the O($p^6$) calculation predicts the 
$K_L\to\pz\ee\g$ branching ratio to be $2.4\times 10^{-8}$. This
prediction depends upon the value of $a_V$. For a value
of $a_V=-0.46$ the branching ratio prediction is $1.51\times 10^{-8}$,
which is consistent with our measurement. The O($p^4$) calculation,
meanwhile, predicts a branching ratio of $1\times 10^{-8}$. Therefore,
our latest measurement disfavors the O($p^4$) prediction while 
favoring the O($p^6$) calculation.

\section{Conclusions}
We have determined the branching ratio B($K_L\to\pz\ee\g$) using
the combined 1997 and 1999 data sets from the KTeV experiment. The statistics
represents a factor of 2.5 over our published 1997 result.
Compared to our previous result, this analysis utilizes a number
of new analysis techniques and employs an improved 
understanding of the backgrounds. We determine the
branching ratio to be
B($K_L\to\pz\ee\g$) = $(1.62\pm 0.14_{stat}\pm 0.09_{syst})\times 10^{-8}$.
The most recent measurement of $a_V$\cite{ref:NA48} suggests that the decay
$K_L\to\pz\ee$ is dominated by a CP violating amplitude. While
the statistics are low, the value of $a_V$ from our determination
is consistent with these conclusions. Our branching ratio measurement
confirms the value of $a_V$ measured by NA48 and indicates that 
O($p^6$) terms are important for modeling this decay mode.
The measured branching ratio is within 0.7 $\sigma$ of the
O($p^6$) prediction and about 3.7 $\sigma$ from the O($p^4$)
prediction.
A factor of twenty increase in
the statistics of $K_L\to\pz\ee\g$ would make the measurement
of $a_V$ competitive with the current best measurements.
Because of the kinematics and small branching ratio for
$K_L\to\pz\ee\g$, this decay will not constitute a large background
in future searches for $K_L\to\pz\ee$.

\begin{acknowledgments}

We gratefully acknowledge the support and effort of the Fermilab
staff and the technical staffs of the participating institutions for
their vital contributions.  This work was supported in part by the U.S.
Department of Energy, The National Science Foundation, The Ministry of
Education and Science of Japan,
Fundação de Amparo a Pesquisa do Estado de S\~ao Paulo-FAPESP,
Conselho Nacional de Desenvolvimento Cientifico e Tecnologico-CNPq and
CAPES-Ministerio Educao.
\end{acknowledgments}


\end{document}

%% file: ktev99_author_list.tex
%

\newcommand{\UAz}{University of Arizona, Tucson, Arizona 85721}
\newcommand{\UCLA}{University of California at Los Angeles, Los Angeles,
                    California 90095} 
\newcommand{\Campinas}{Universidade Estadual de Campinas, Campinas, 
                       Brazil 13083-970}
\newcommand{\EFI}{The Enrico Fermi Institute, The University of Chicago, 
                  Chicago, Illinois 60637}
\newcommand{\UB}{University of Colorado, Boulder, Colorado 80309}
\newcommand{\ELM}{Elmhurst College, Elmhurst, Illinois 60126}
\newcommand{\FNAL}{Fermi National Accelerator Laboratory, 
                   Batavia, Illinois 60510}
\newcommand{\Osaka}{Osaka University, Toyonaka, Osaka 560-0043 Japan} 
\newcommand{\Rice}{Rice University, Houston, Texas 77005}
\newcommand{\SaoPaolo}{Universidade de S\~ao Paulo, S\~ao Paulo, Brazil 05315-970}
\newcommand{\UVa}{The Department of Physics and Institute of Nuclear and 
                  Particle Physics, University of Virginia, 
                  Charlottesville, Virginia 22901}
\newcommand{\UW}{University of Wisconsin, Madison, Wisconsin 53706}

\affiliation{\UAz}
\affiliation{\UCLA}
\affiliation{\Campinas}
\affiliation{\EFI}
\affiliation{\UB}
\affiliation{\ELM}
\affiliation{\FNAL}
\affiliation{\Osaka}
\affiliation{\Rice}
\affiliation{\SaoPaolo}
\affiliation{\UVa}
\affiliation{\UW}

\author{E.~Abouzaid}	  \affiliation{\EFI}
\author{M.~Arenton}       \affiliation{\UVa}
\author{A.R.~Barker}      \altaffiliation[Deceased.]{ } \affiliation{\UB}
\author{L.~Bellantoni}    \affiliation{\FNAL}
\author{E.~Blucher}       \affiliation{\EFI}
\author{G.J.~Bock}        \affiliation{\FNAL}
\author{E.~Cheu}          \altaffiliation[To whom correspondence should be
                          addressed. Electronic address: 
                          elliott@physics.arizona.edu]{} \affiliation{\UAz}
\author{R.~Coleman}       \affiliation{\FNAL}

\author{B.~Cox}           \affiliation{\UVa}
\author{A.R.~Erwin}       \affiliation{\UW}
\author{C.O.~Escobar}     \affiliation{\Campinas}  
\author{A.~Glazov}        \affiliation{\EFI}
\author{A.~Golossanov}    \affiliation{\UVa} 
\author{R.A.~Gomes}       \affiliation{\Campinas}
\author{P. Gouffon}       \affiliation{\SaoPaolo}
\author{Y.B.~Hsiung}      \affiliation{\FNAL}
\author{D.A.~Jensen}      \affiliation{\FNAL}
\author{R.~Kessler}       \affiliation{\EFI}
\author{Y.J.~Kim}         \affiliation{\UAz}
\author{K.~Kotera}	  \affiliation{\Osaka}
\author{A.~Ledovskoy}     \affiliation{\UVa}
\author{P.L.~McBride}     \affiliation{\FNAL}

\author{E.~Monnier}
   \altaffiliation[Permanent address ]{C.P.P. Marseille/C.N.R.S., France}
   \affiliation{\EFI}  

\author{K.S.~Nelson}     \affiliation{\UVa}  
\author{H.~Nguyen}       \affiliation{\FNAL}
\author{R.~Niclasen}     \affiliation{\UB}
\author{D.G.~Phillips~II} \affiliation{\UVa}
\author{H.~Ping}         \affiliation{\UW}  
\author{E.J.~Ramberg}    \affiliation{\FNAL}
\author{R.E.~Ray}        \affiliation{\FNAL}
\author{M.~Ronquest}     \affiliation{\UVa}
\author{E.~Santos}       \affiliation{\SaoPaolo}
\author{W.~Slater}       \affiliation{\UCLA}
\author{D.~Smith}        \affiliation{\UVa}
\author{N.~Solomey}      \affiliation{\EFI}
\author{E.C.~Swallow}    \affiliation{\EFI}\affiliation{\ELM}
\author{P.A.~Toale}      \affiliation{\UB}
\author{R.~Tschirhart}   \affiliation{\FNAL}
\author{C.~Velissaris}   \affiliation{\UW}  
\author{Y.W.~Wah}        \affiliation{\EFI}
\author{J.~Wang}         \affiliation{\UAz}
\author{H.B.~White}      \affiliation{\FNAL}
\author{J.~Whitmore}     \affiliation{\FNAL}
\author{M.~J.~Wilking}      \affiliation{\UB}
\author{R.~Winston}      \affiliation{\EFI}
\author{E.T.~Worcester}  \affiliation{\EFI}
\author{M.~Worcester}    \affiliation{\EFI}
\author{T.~Yamanaka}     \affiliation{\Osaka}
\author{E.~D.~Zimmerman} \affiliation{\UB}
\author{R.F.~Zukanovich} \affiliation{\SaoPaolo}